\documentclass[a4paper]{jpconf}
\usepackage{graphicx,natbib,astro,url}
\begin{document}
\title{Imaging black holes: past, present and future}

\author{Heino Falcke$^{1,2}$}

\address{$^1$ Department of Astrophysics/IMAPP, Radboud University, P.O. Box 9010, 6500 GL Nijmegen, The Netherlands}
\address{$^2$ ASTRON, Dwingeloo, The Netherlands}
\ead{h.falcke@astro.ru.nl}
\address{}
\address{Proceedings of the 3rd Karl Schwarzschild Meeting, Journal of Physics Conference Series Vol 942, conference 1, \url{http://iopscience.iop.org/article/10.1088/1742-6596/942/1/012001}}

\begin{abstract}
  This paper briefly reviews past, current, and future efforts to
  image black holes.  Black holes seem like mystical
  objects, but they are an integral part of current astrophysics and
  are at the center of attempts to unify quantum physics and general
  relativity. Yet, nobody has ever seen a black hole. What do they look
  like? Initially, this question seemed more of an academic
  nature. However, this has changed over the past two
  decades. Observations and theoretical considerations suggest that
  the supermassive black hole, Sgr~A*, in the center of our Milky Way
  is surrounded by a compact, foggy emission region radiating at and
  above 230 GHz. It has been predicted that the event horizon of
  Sgr~A* should cast its shadow onto that emission region, which could
  be detectable with a global VLBI array of radio telescopes. In
  contrast to earlier pictures of black holes, that dark feature is
  not supposed to be due to a hole in the accretion flow, but would
  represent a true negative image of the event horizon. Currently, the
  global Event Horizon Telescope consortium is attempting to
  make such an image.  In the future those images could be improved by
  adding more telescopes to the array, in particular at high sites in
  Africa. Ultimately, a space array at THz frequencies, the Event
  Horizon Imager, could produce much more detailed images of black
  holes. In combination with numerical simulations and precise
  measurements of the orbits of stars -- ideally also of pulsars --
  these images will allow us to study black holes with unprecedented
  precision.
\end{abstract}

\section{Introduction}
When the name black hole was coined in the sixties -- allegedly a term
picked up by science journalist Ann Ewing and later popularized by
John
A. Wheeler\footnote{\url{http://www.worldwidewords.org/topicalwords/tw-bla1.htm}}
-- it sparked the imagination: not a dull mathematical, abstract
object in a theory of gravity but a somewhat mythical, perhaps scary
object with an intriguing visual appearance --- something seemingly
invisible that nonetheless exists. Ever since, there have been many
depictions of black holes in the scientific literature, in the arts,
and in the public media --- perhaps even more so in the latter.

Ironically, the reason why black holes became fashionable was not
because they were black, but because they were bright.  This was related to the
discovery of quasars more than 50 years ago \citep[see,
  e.g.][]{SulenticMarzianiDOnofrio2012a}. Quasars are
extremely luminous and simple arguments could show that the deep
potential well of black holes could in principle
produce that energy.

A falling test mass $m$ at radius $R$ from a black hole mass
$M_\bullet$ woul have a Newtonian potential of $E=G M_\bullet m/R$,
where $G$ is the gravitational constant. For a constant mass inflow
rate $\dot m$ the available energy per time $\dot E$ can be converted
into a luminosity $L$ with an assumed efficiency $\eta$ of order
$10\%$ to yield $ L \sim \eta \dot E=\eta {G M_\bullet\over R} \dot m
={1\over2} \eta \left({R_{\rm s}\over R}\right)\dot m c^2$, where $c$
is the speed of light and $R$ is assumed to be of order the
Schwarzschild radius, $R_{\rm S}=2 GM_\bullet c^{-2}$. For $\dot m
\sim 1 M_\odot/$yr the black hole luminosity $L$ can be of order
$10^{46}$ erg/sec for black holes some hundred million solar masses --
comparable to inferred quasar luminosities.

Not surprisingly, the topic of early research on black holes and
active galactic nuclei (AGN) then mainly focused on the overall
spectral energy distribution (SED) of luminous AGN. Because of their
large distances, imaging a black hole on scales of the event horizon
seemed out of reach and more like a purely academic exercise. The only
resolvable structure of quasars was seen in their radio emission that
later was identified as coming from relativistic plasma jets emerging
from near the putative black holes \citep[e.g.,][]{BridlePerley1984,Ferrari1998}.

\begin{figure}[h]
\includegraphics[width=\textwidth]{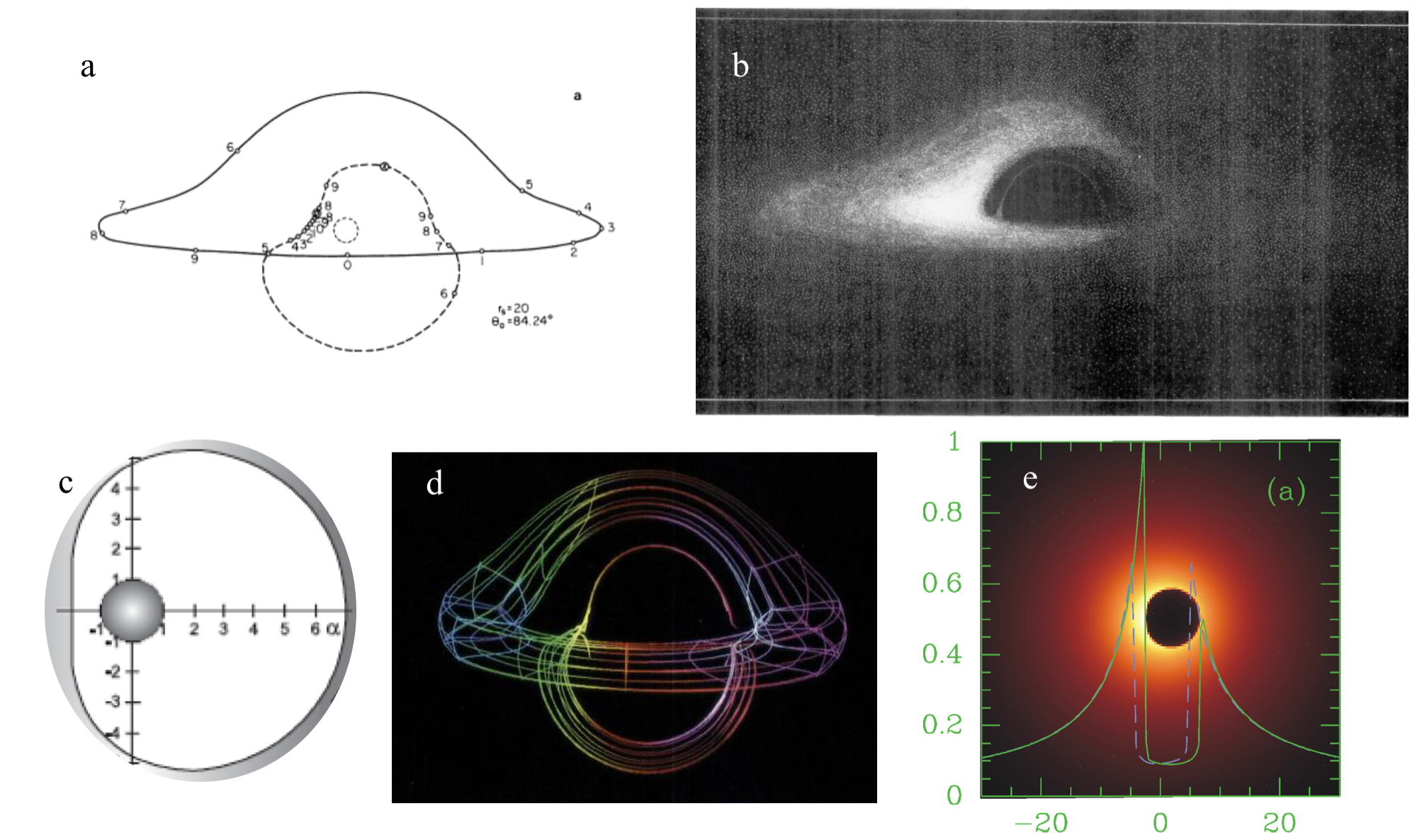}\hspace{2pc}%
\begin{minipage}[b]{\textwidth}\caption{\label{fig-earlyimages}
    Early representations of black holes. From top left to bottom
    right: a) star orbiting a black hole \citep{CunninghamBardeen1973a},
    b) thin accretion disk extending down to the innermost stable circular orbit (ISCO) \citep{Luminet1979a},
    c) an extended stellar disk far behind a black hole (Bardeen 1973, as reproduced in Falcke
    1999\nocite{Falcke1999b}), d) inner edge of an accretion disk seen with
    color \citep{Viergutz1993a}, and e) shadow of a black hole
    surrounded by an optically thin emission region \citep{FalckeMeliaAgol2000}.}
\end{minipage}
 \end{figure}

\section{Black hole images: early predictions}
So, what does a black hole really look like? The first attempt I am
aware of\footnote{Cunningham \& Bardeen also refer to
  \citet{CampbellMatzner1973a}, doing a similar calculation for a
  Schwarzschild black hole, but neither motivation nor results seem particularly relevant here.} dates back
to \cite{CunninghamBardeen1973a}, where the authors calculate the
appearance of a star in a tight orbit (3 and 20 Schwarzschild radii)
around a Kerr black hole (Fig.~\ref{fig-earlyimages}a). The apparent
orbit already traces out the basic structures later also seen by more
complete accretion disk simulations. However, what we see in these
images is not an image of the event horizon at all, it is actually the
lensed image of the stellar orbit and true for any gravitating object
with or without an event horizon.

The paper that actually caught my attention first on this topic was by
\cite{Bardeen1973}, published in a conference proceedings, where it seemed
hardly noticed and cited until then\footnote{Intriguingly, this is the book,
  where the famous paper by \citet{NovikovThorne1973a} was published
  and which has been cited more than 500 times by now.}. In this paper Bardeen
calculates the appearance of an extended stellar disk behind a Kerr
black hole (Fig.~\ref{fig-earlyimages}c). A similar calculation was
later presented by \cite{Luminet1979a} for an extended plane wave
source at a large distance, e.g., a star in a binary system or simply
a large flash light shining at the black hole from behind.

Here an important effect is noticeable, which is related to the photon
orbit. The photon orbit is an orbit where light can go around black
holes on a closed loop. In practice this means that light entering at
or around the photon orbit, will experience a sharp divide: if the
light ray is a little bit too close it will disappear behind the event
horizon, if it is a little bit outside, it will be able to escape to
infinity. The consequence is indeed a sharp image of the event
horizon.

In the same paper Luminet also presents his famous image
(Fig.~\ref{fig-earlyimages}b). The figure is hailed as the first ever
computer-generated image of a black hole. However, as in the case of
\cite{CunninghamBardeen1973a}, the dark structure traced out in the
center is not actually an image of the black hole itself, but is a lensed
image of the inner edge of the accretion disk. Consequently, the image
depends crucially on the assumption of where the disk ends. The
standard view at the time \citep{NovikovThorne1973a,PageThorne1974}
was that the disk terminated and stopped radiating at the innermost stable circular radius
(ISCO, then often called marginally stable radius $r_{\rm ms}$). The
same is true for later efforts
\citep[][Fig.~\ref{fig-earlyimages}d]{Viergutz1993a}. Strictly
speaking we see therefore in these images mainly a hole in an
accretion disk rather than the black hole itself.

\section{The Galactic Center: where dreams come true}
A region of high hopes for black hole research had been the center of
our own Galaxy \citep{Lynden-BellRees1971}. The compact radio source,
Sgr~A*, found by \citet{BalickBrown1974a}, was often compared to radio
cores in AGN, and hence considered to mark the supermassive black hole
in our Milky Way. Near-infrared measurements of stellar orbits \citep{EckartGenzel1996,GhezKleinMorris1998}, revealing
stars moving with up to 10,000 km/s around this radio source, and
very-long baseline interferometry (VLBI) observations
\citep{ReidBrunthaler2004a}, have confirmed that indeed there is a dark
mass of about 4.3 million solar masses
\citep{GillessenPlewaEisenhauer2017a} within a few Schwarzschild radii
of Sgr~A*
\citep[see][for
  reviews]{GenzelEisenhauerGillessen2010a,FalckeMarkoff2013a}\footnote{For
  parameters of Sgr~A* used here, see the overview in Table 1 in
  Falcke \& Markoff (2013).}.

The nature of this radio emission was long under debate
\citep{MeliaFalcke2001a}. In the early nineties, we developed a model to
describe Sgr~A* as a starving black hole with a low-power radio jet,
fed by a highly sub-Eddington accretion flow
\citep{FalckeMannheimBiermann1993a,FalckeMarkoff2000a}. A competing
model was to explain all the radio emission from a radiatively
inefficient accretion flow (RIAF) \citep{NarayanYiMahadevan1995a} at
much higher accretion rates. Today we often combine those elements
\citep{YuanMarkoffFalcke2002,MoscibrodzkaFalcke2013a} and
keep the low accretion rate now demanded by polarization
observations of Sgr~A*
\citep{BowerWrightFalcke2003a,MarroneMoranZhao2007}.

A prediction of the jet model was that the radio emission should come
from closer to the black hole the higher the frequency of the emission
was, meaning that the submm-wave bump \citep{MezgerZylkaSalter1989} in
the spectrum should come from a compact emission region near the event
horizon. One of the first multi-telescope and multi-wavelength
campaigns for Sgr~A*, involving four telescopes, confirmed that
picture \citep{FalckeGossMatsuo1998}. Simple theory also predicted
that the emission should become optically thin at and above 230
GHz. Hence, we proposed in the same paper that one should be able to
to image the event horizon against this background with submm-wave VLBI
observations.

Since the original Bardeen picture only considered emission from far
behind the black hole, we then calculated the appearance of a black
hole surrounded by an optically thin emission region
\citep[][Fig.~\ref{fig-earlyimages}e]{FalckeMeliaAgol2000}, which was
more applicable to Sgr~A*. Indeed, we found a dark region of about
some 5 Schwarzschild radii in diameter, surrounded by a bright ring of
emission. That dark region is not just a sharp silhouette as sometimes
implied. Interstellar scattering of radio waves and a blend of
emission in front of and behind the black hole will give the feature a
more diffuse appearance. Hence, we called it the ``shadow'' of the
event horizon. Many papers now exist that calculate black holes
shadows, also in non-standard theories of gravity, suggesting its usefulness for testing GR \citep[see][for
  reviews]{FalckeMarkoff2013a,Johannsen2016b}.

Simulations are now becoming much more sophisticated and are done
using numerical general-relativistic magnetohydrodynamic simulations
(GRMHD) with ray tracing to take the astrophysical model into account.
This was pioneered by \cite{MoscibrodzkaGammieDolence2009}, initially
for a pure disk model, but now also including jet emission
\citep[see][for a recent review]{Moscibrodzka2017a}. The new BHAC
code \citep{PorthOlivaresMizuno2017a} now also allows one to perform
these simulations in arbitrary parametrized space times
\citep{KonoplyaRezzollaZhidenko2016a}. In
\citet{MoscibrodzkaFalcke2013a} and \citet{MoscibrodzkaFalckeShiokawa2014a} we
showed that the original simple analytical jet model does in fact hold
up in GRMHD simulations, reproduces the observational characteristics,
and produces radio mm-wave emission close to the event horizon ---
just as needed for shadow imaging.

The diameter of the shadow would be of order $10\,G M_\bullet c^{-2}$,
i.e.~63 million km for $M_\bullet=4.3\times10^6 M_\odot$,
corresponding to 51 microarcseconds ($\mu$as) for a distance of 8.3
kpc to the Galactic Center. The resolution of a radio interferometer
is given by the ratio of observing wavelength and telescope separation
$\lambda/D$. For global interferometers $D$ reaches typical values of
15,000 to 6,000 km, giving one a resolution between 18 to 50 $\mu$as
at 230 GHz (i.e., $\lambda1.3$mm) --- just large enough to start
resolving the shadow. Moreover, the scattering effect decreases with
$\lambda^2$, also approaching 22 $\mu$as at $\lambda=1.3$ mm. Hence,
230 GHz becomes a `magical' observing frequency, where interstellar
scattering, optical depth in the source, global VLBI resolution, and
atmospheric transmission make this experiment just possible. At higher
frequencies the earth's atmosphere becomes too hostile and at lower
frequencies the source is scatter broadened and intransparent. Hence,
we can thank God, that the earth is just large enough and at the right
location in our Galaxy, to let us to not only see the light, but also
the shadow of a black hole.

\section{The experimental development: from the past to the present}

The first successful 1.3 mm VLBI experiment measuring Sgr~A* was a
single baseline observation between the IRAM telescopes on Pico Veleta
(Spain) and Plateau de Bure (France) led by the MPIfR in Bonn
\citep{KrichbaumGrahamWitzel1998a}\footnote{The observation was at low
  signal-to-noise and the size seemed much larger than the event
  horizon, hence, did not get too much attention. However, at the time
  the black hole mass was thought to be much smaller than it is now
  and the size expressed in Schwarzschild radii seemed larger than the
  shadow size. In fact, the measured size in $\mu$as at 230 GHz is
  consistent with current measurements within the error bars.}.  The
idea of imaging the event horizon in Sgr~A* was first more widely
discussed -- though somewhat skeptically -- at the Galactic Center
conference in Tucson 1998 \citep[e.g., see transcript of VLBI
  discussions in][]{ZensusFalcke1999a}, followed by more discussions
at other meetings. In 2004, we organized an informal evening session
at the 30th anniversary meeting on the occasion of the discovery of
Sgr~A*\footnote{\url{http://www.aoc.nrao.edu/~gcnews/GCconfs/SgrAstar30} --- see small print
  after ``Poster Contributions''.}, where the community was finally
ready to embrace and support such an experiment\footnote{In fact,
  expressed in the form of a vote and a signed declaration.}. The
three speakers at the meeting, G. Bower, S. Doeleman and myself,
started a series of telecons, further developing the idea of a global
physics-like collaboration for black hole imaging. MIT Haystack
observatory kept improving digital VLBI hardware for 3 and 1mm
observing and proceeded further in a small ad hoc collaboration
\citep{DoelemanWeintroubRogers2008a}. A little earlier, regular 3mm
observing had already been made possible via a joint memorandum of
understanding between multiple observatories, forming the Global
mm-VLBI Array (GMVA, 2003), coordinated by the MPIfR Bonn.

  Significant progress was made at lower
  frequencies. \citet{BowerFalckeHerrnstein2004} were able to measure
  the intrinsic size at 7mm and 13 mm, showing for the first time that
  indeed Sgr~A* decreases in size towards higher frequencies as
  predicted and that it would approach event horizon scales if
  extrapolated. This trend was soon confirmed by better VLBI
  measurements at 3mm \citep{ShenLoLiang2005}.

  \citet{DoelemanWeintroubRogers2008a} published the first
  three-baseline experiment, demonstrating a source size of
  $37^{+16}_{-10}\;\mu$as, i.e.~smaller than the shadow size. Within
  the errors the results were consistent with the
  \citet{KrichbaumGrahamWitzel1998a} results, but more robust and with
  significantly smaller error bars. The triangle SMTO (Arizona) --
  CARMA (California) -- JCMT/SMA (Hawaii) had a number of productive
  observing runs, showing asymmetric structure and evidence for
  homogenous magnetic fields
  \citep{FishJohnsonDoeleman2016a,JohnsonFishDoeleman2015a}. Also, a
  four-baseline experiment, involving the APEX telescope, succeeded,
  revealing evidence for more event-horizon scale structure
  \citep[][Lu et al. 2018, in prep.]{KrichbaumRoyLu2014a}.

  Community white papers to support mm-VLBI efforts were published in
  the US, Europe, and Asia
  \citep{FalckeLaingTesti2012a,FishAlefAnderson2013a,TilanusKrichbaumZensus2014a,AsadaKinoHonma2017a}
  and mm-VLBI was included in the European Science Vision for
  astronomy\footnote{\url{http://www.eso.org/public/news/eso0744/}} in
  2007 and in the US Astro2010 decadal review.  In the summer of 2017
  a formal memorandum of understanding was signed by 13 stakeholder
  institutions with about 150 individual scientists to form the Event
  Horizon Telescope Consortium (EHTC), which currently is conducting
  observations with 8 telescopes (IRAM 30m, JCMT, SMA, SMTO, SPT, LMT,
  ALMA, APEX) on six mountains\footnote{The term ``Event Horizon
    Telescope'' came out of a coffee table discussion at the AAS
    meeting in Long Beach in 2009.}. Besides outfitting telescopes
  with proper VLBI-equipment, getting ALMA VLBI-ready had been another
  important step \citep{MatthewsCrewDoeleman2017a}. The EHTC experiments are ongoing and one has to
  await further data analysis, before any results can be discussed.
  
  Despite the high bandwidth and some large telescopes, the limited
  number of simultaneous baselines will make imaging still
  difficult. Simulations by \citet{LuRoelofsFish2016a} suggest that
  images made during a single day suffer significant uncertainties
  from source variability. Substructures and variations in the
  scattering screen will amplify this uncertatiny further. Hence,
  multiple observing epochs may need to be added for a good image.

  In the end, one will have to fit astrophysical shadow images to the
  experimental data \citep[e.g.,][]{BroderickJohannsenLoeb2014a} to
  derive black hole parameters and to verify GR. However, the variations
  in those models lead to an unknown systematic error that still needs to be
  quantified better. Assuming GR is correct, shadow imaging could lead
  to an accuracy of up to 10 \% \citep{PsaltisOzelChan2015a} in confirming
  the null-hypothesis that a GR event horizon exists. If, however, GR
  itself is to be tested against alternative models, it may not be
  that easy to discriminate between GR shadow images and non-GR images
  \citep{MizunoDilaton2018}.

  Hence, like in cosmology, where multiple experiments are combined to
  provide, e.g., constraints on the nature of dark energy, it may also
  become important to add complementary information to the VLBI
  data. Psaltis, Kramer, \& Wex (2016) \nocite{PsaltisWexKramer2016a}
  show that measuring a pulsar in the Galactic center would be ideal
  for this. Indeed, after decades of searches, a first pulsar has been
  found \citep{EatoughFalckeKaruppusamy2013a} but closer and older
  ones are needed to measure the spacetime around Sgr~A* accurately.
  The SKA, but also ALMA, could detect more pulsars in the
  future. Moreover, more and better stellar orbits will also constrain
  spacetime in a significant way. Here ESO's new near-infrared
  interferometer, GRAVITY, will make a big impact
  \citep{Gravity-CollaborationAbuterAccardo2017a}.

  \section{The future: Africa mm-wave telescope and space VLBI}
  The global millimeter VLBI array will further expand. ASIAA plans to
  add the Greenland telescope (GLT) for observations of M87
  \citep{MatsushitaAsadaBlundell2015a} and the BlackHoleCam team
  \citep{GoddiFalckeKramer2017a} together with IRAM, is preparing to
  phase up\footnote{Phasing-up an array means to add the signals from
    individual antennas in an array in phase, such that the telescopes
    act collectively as a large single dish.} the NOEMA array (Plateau
  de Bure, France) for VLBI. Also, the University of Arizona Kitt Peak
  telescope and the LLAMA dish (Brazil, Argentina) may be added soon.
  
\begin{figure}[htb]
\includegraphics[width=\textwidth]{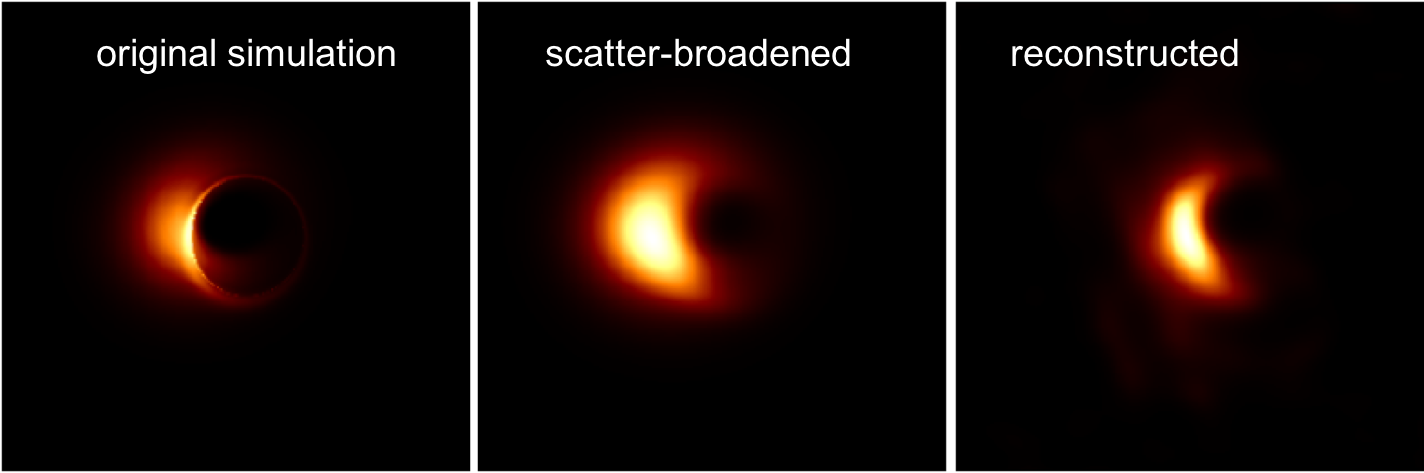}\hspace{2pc}%
\begin{minipage}[b]{\textwidth}\caption{\label{figAMT}Left: Average image of the time variable black hole ``movie'' used in Lu et al. (2016), b) same image convolved with scattering kernel to account for interstellar scattering, c) synthetic image recovered with an array including SMA, SMT, LMT, ALMA, IRAM 30m, NOEMA, SPT, and an Africa telescope (AMT) using the procedure described in Lu et al. and averaging 8 epochs.}
\end{minipage}
 \end{figure}

  Moreover, we are currently investigating possibilities to add an
  African mm-wave telescope (AMT) to the array. The Gamsberg mountain
  in Namibia is a potential site
  \citep{BackesMullerConway2016a}. First simulations
  \citep[][see also Fig.~\ref{figAMT}]{RoelofsEHI2017} indicate that an Africa
  telescope would add more robustness and quality to the imaging. If
  multiple epochs are added, the AMT -- despite being smaller -- would
  become as important as ALMA for the image quality. In the case of
  long observations and high signal-to-noise ratio, the distribution
  of the telescopes is more important than the sensitivity of
  individual telescopes. Of course, more telescopes in Africa, e.g.,
  near Kilimanjaro in Tanzania, or on the Drakensberg in Lesotho,
  would improve the quality even further.

\begin{figure}[htb]
\includegraphics[width=\textwidth]{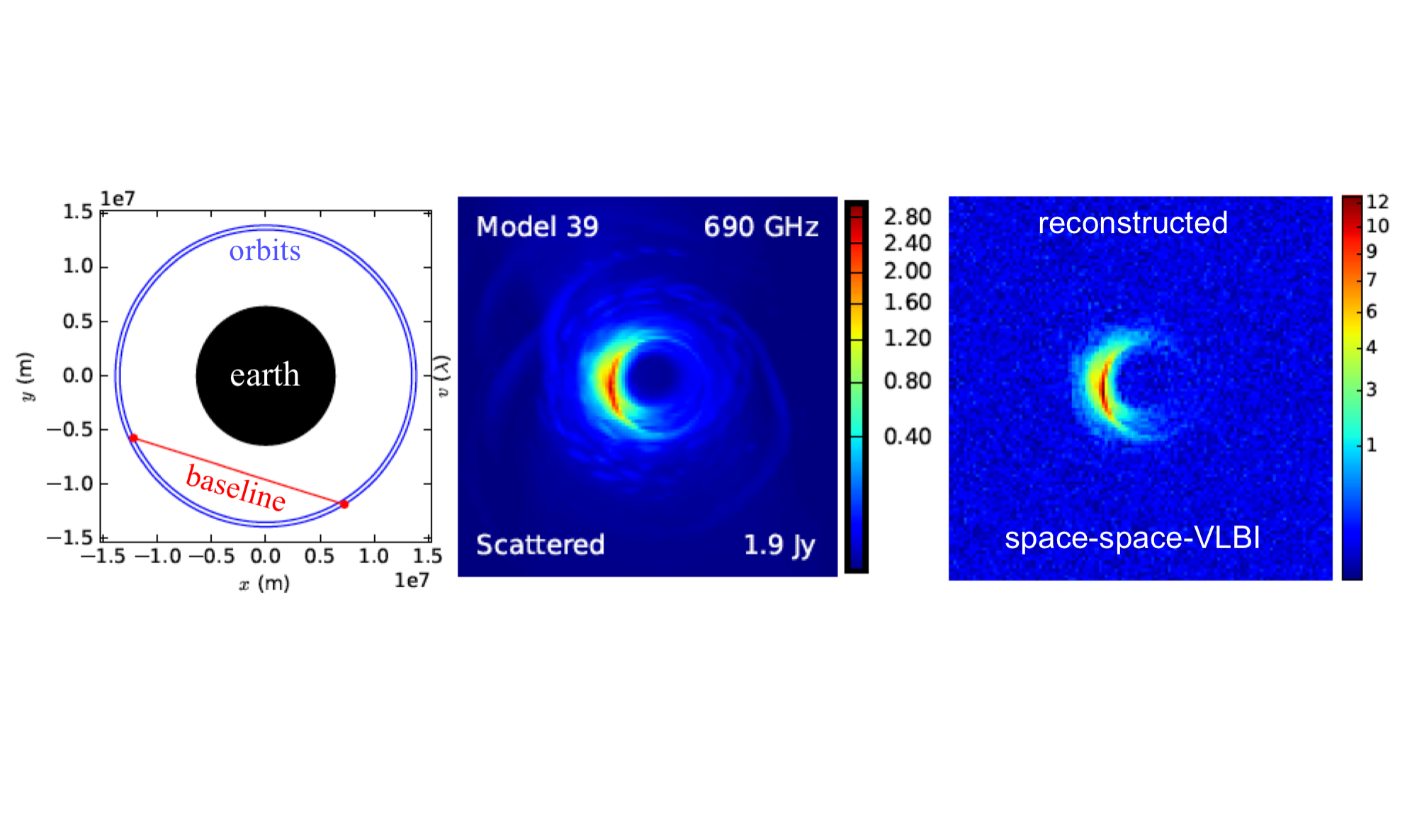}\hspace{2pc}%
\begin{minipage}[b]{\textwidth}\caption{\label{figEHI}Simulated Space VLBI (EHI) image from \citet{RoelofsEHI2017}. Left: Schematic of the orbits (blue) of the two spacecraft (red points), Middle: Model image at 690 GHz from \citet{MoscibrodzkaFalckeShiokawa2014a}, Right: images recovered from synthetic image data integrated over 2 years. The synthetic data were generated using the EHT imaging software from \citet{ChaelJohnsonNarayan2016a}.}
\end{minipage}
 \end{figure}

Of course, as always, space is the final
frontier. \citet{RoelofsEHI2017} performed simulations using a
two-element space-to-space interferometer, the Event Horizon Imager
(EHI), consisting of two spacecraft at about 15,000 km. The telescopes
are separated by some 50 km in altitude (Fig.~\ref{figEHI}) and
observe at 690 GHz. At these frequencies interstellar scattering is
almost negligible ($\sim2\mu$as), the source will be very optically
thin, and a ground-based VLBI array is unfeasible due to strong and
variable tropospheric absorption. Having only a single space-to-space
baseline, rather than a fleet of spacecraft, greatly reduces the
complexity.

While orbiting the earth, the spacecraft slowly drift apart due to
different orbital speeds. Eventually, the connecting baseline will
have assumed all orientations and separations, giving one a spiral
pattern in the uv-plane --- the Fourier plane of the image. A filled
uv-plane greatly improves image reconstruction in interferometric
measurements. For realistic receiver noise and source fluxes, two 6~m
dishes could obtain a very detailed picture of the black hole shadow
(Fig.~\ref{figEHI}) with 4 $\mu$as resolution, when integrating over
several months or even years --- thereby also integrating over source
variability. The constant features are those produced by GR.

\section{Conclusion}
All theoretical and experimental data support the idea that imaging
the shadow of the black hole in the Galactic Center (and probably also
in M87) should be possible with global radio interferometers at 230
GHz and higher. For testing general relativity, it is crucial that the
emission region indeed extends well inside the ISCO, otherwise we
would mainly see a hole in the accretion flow, rather than a black
hole shadow. Hence, understanding the astrophysical model remains
important. Current GRMHD models with a radiatively inefficient
accretion flow and a jet outflow already explain the data for Sgr~A*
and M87* well. The newly formed Event Horizon Telescope Consortium is
now attempting to make the first such images. Simulations suggest that
multiple epochs may be necessary to obtain a robust image. If
successful they could verify the prediction of an event horizon to
within 10\% accuracy, assuming GR is correct. Testing alternative
theories of gravity would benefit from additional inputs, such as
better stellar orbits and ideally, a pulsar around the black
hole. Images can be improved by adding additional telescopes to the
global array. An African mm-wave telescope (AMT) would add crucial new
baselines, producing better and more robust images. Ultimately, a
space interferometer, such as the Event Horizon Imager operating at
THz frequencies, could produce a quantum leap in black hole
imaging. This may be possible even with just two spacecraft. Hence,
it is probably more a question of when, rather than whether, the first
good images of black holes will arrive. Since most supermassive black
holes stay around for a long time, those images can and will become
better with time. This provides us with the golden opportunity to
conduct more and more detailed tests of gravity in its most extreme
limit.

\ack I am grateful for comments from F. Roelofs, M. Moscibrodzka,
A. Roy, N. Nagar, G. Crew, and other colleagues. The author is supported by the European
Research Council (ERC) Synergy Grant “BlackHoleCam” (Grant 610058),
which is jointly led by the author togehther with L. Rezzolla
(Univ. Frankfurt), and M. Kramer (MPIfR Bonn).




\begin{thebibliography}{61}
\expandafter\ifx\csname natexlab\endcsname\relax\def\natexlab#1{#1}\fi

\bibitem[{{Asada} {et~al.}(2017){Asada}, {Kino}, {Honma}, {Hirota}, {Lu},
  {Inoue}, {Sohn}, {Shen}, {Ho}, {Akiyama}, {Algaba}, {An}, {Bower}, {Byun},
  {Dodson}, {Doi}, {Edwards}, {Fujisawa}, {Gu}, {Hada}, {Hagiwara},
  {Jaroenjittichai}, {Jung}, {Kawashima}, {Koyama}, {Lee}, {Matsushita},
  {Nagai}, {Nakamura}, {Niinuma}, {Phillips}, {Park}, {Pu}, {Ro}, {Stevens},
  {Trippe}, {Wajima}, \& {Zhao}}]{AsadaKinoHonma2017a}
{Asada}, K., {Kino}, M., {Honma}, M., {et~al.} 2017, ArXiv e-prints, 1705.04776

\bibitem[{{Backes} {et~al.}(2016){Backes}, {M\"uller}, {Conway}, {Deane},
  {Evans}, {Falcke}, {Fraga-Encinas}, {Goddi}, {Klein Wolt}, {Krichbaum},
  {MacLeod}, {Ribeiro}, {Roelofs}, {Shen}, \& {van
  Langevelde}}]{BackesMullerConway2016a}
{Backes}, M., {M\"uller}, C., {Conway}, J.~E., {et~al.} 2016, in Proceedings of
  the 4th Annual Conference on High Energy Astrophysics in Southern Africa
  (HEASA 2016). January 13th, 2016. South African Astronomical Observatory
  (SAAO), Cape Town, South Africa., 29

\bibitem[{{Balick} \& {Brown}(1974)}]{BalickBrown1974a}
{Balick}, B. \& {Brown}, R.~L. 1974, \apj, 194, 265

\bibitem[{Bardeen(1973)}]{Bardeen1973}
Bardeen, J.~M. 1973, in Black Holes, ed. {C. DeWitt} \& {B. S. DeWitt} (New
  York: Gordon \& Breach), 215--239

\bibitem[{{Bower} {et~al.}(2004){Bower}, {Falcke}, {Herrnstein}, {Zhao},
  {Goss}, \& {Backer}}]{BowerFalckeHerrnstein2004}
{Bower}, G.~C., {Falcke}, H., {Herrnstein}, R.~M., {et~al.} 2004, Science, 304,
  704

\bibitem[{{Bower} {et~al.}(2003){Bower}, {Wright}, {Falcke}, \&
  {Backer}}]{BowerWrightFalcke2003a}
{Bower}, G.~C., {Wright}, M.~C.~H., {Falcke}, H., \& {Backer}, D.~C. 2003,
  \apj, 588, 331

\bibitem[{{Bridle} \& {Perley}(1984)}]{BridlePerley1984}
{Bridle}, A.~H. \& {Perley}, R.~A. 1984, \araa, 22, 319

\bibitem[{{Broderick} {et~al.}(2014){Broderick}, {Johannsen}, {Loeb}, \&
  {Psaltis}}]{BroderickJohannsenLoeb2014a}
{Broderick}, A.~E., {Johannsen}, T., {Loeb}, A., \& {Psaltis}, D. 2014, \apj,
  784, 7

\bibitem[{{Campbell} \& {Matzner}(1973)}]{CampbellMatzner1973a}
{Campbell}, G.~A. \& {Matzner}, R.~A. 1973, Journal of Mathematical Physics,
  14, 1

\bibitem[{{Chael} {et~al.}(2016){Chael}, {Johnson}, {Narayan}, {Doeleman},
  {Wardle}, \& {Bouman}}]{ChaelJohnsonNarayan2016a}
{Chael}, A.~A., {Johnson}, M.~D., {Narayan}, R., {et~al.} 2016, \apj, 829, 11

\bibitem[{{Cunningham} \& {Bardeen}(1973)}]{CunninghamBardeen1973a}
{Cunningham}, C.~T. \& {Bardeen}, J.~M. 1973, \apj, 183, 237

\bibitem[{{Doeleman} {et~al.}(2008){Doeleman}, {Weintroub}, {Rogers},
  {Plambeck}, {Freund}, {Tilanus}, {Friberg}, {Ziurys}, {Moran}, {Corey},
  {Young}, {Smythe}, {Titus}, {Marrone}, {Cappallo}, {Bock}, {Bower},
  {Chamberlin}, {Davis}, {Krichbaum}, {Lamb}, {Maness}, {Niell}, {Roy},
  {Strittmatter}, {Werthimer}, {Whitney}, \&
  {Woody}}]{DoelemanWeintroubRogers2008a}
{Doeleman}, S.~S., {Weintroub}, J., {Rogers}, A.~E.~E., {et~al.} 2008, \nat,
  455, 78

\bibitem[{Eatough {et~al.}(2013)Eatough, {H. Falcke}, Karuppusamy, \& {et
  al.}}]{EatoughFalckeKaruppusamy2013a}
Eatough, R.~P., {H. Falcke}, Karuppusamy, R., \& {et al.} 2013, \nat,
  doi:10.1038/nature12499

\bibitem[{{Eckart} \& {Genzel}(1996)}]{EckartGenzel1996}
{Eckart}, A. \& {Genzel}, R. 1996, \nat, 383, 415

\bibitem[{{Falcke}(1999)}]{Falcke1999b}
{Falcke}, H. 1999, in ASP Conf. Ser. 186: The Central Parsecs of the Galaxy,
  ed. H.~{Falcke}, A.~{Cotera}, W.~{Duschl}, F.~{Melia}, \& M.~J. {Rieke} (San
  Francisco: Astronomical Society of the Pacific), 148

\bibitem[{{Falcke} {et~al.}(1998){Falcke}, {Goss}, {Matsuo}, {Teuben}, {Zhao},
  \& {Zylka}}]{FalckeGossMatsuo1998}
{Falcke}, H., {Goss}, W.~M., {Matsuo}, H., {et~al.} 1998, \apj, 499, 731

\bibitem[{{Falcke} {et~al.}(2012){Falcke}, {Laing}, {Testi}, \&
  {Zensus}}]{FalckeLaingTesti2012a}
{Falcke}, H., {Laing}, R., {Testi}, L., \& {Zensus}, A. 2012, The Messenger,
  149, 50

\bibitem[{{Falcke} {et~al.}(1993){Falcke}, {Mannheim}, \&
  {Biermann}}]{FalckeMannheimBiermann1993a}
{Falcke}, H., {Mannheim}, K., \& {Biermann}, P.~L. 1993, \aap, 278, L1

\bibitem[{{Falcke} \& {Markoff}(2000)}]{FalckeMarkoff2000a}
{Falcke}, H. \& {Markoff}, S. 2000, \aap, 362, 113

\bibitem[{{Falcke} \& {Markoff}(2013)}]{FalckeMarkoff2013a}
{Falcke}, H. \& {Markoff}, S.~B. 2013, Classical and Quantum Gravity, 30,
  244003

\bibitem[{{Falcke} {et~al.}(2000){Falcke}, {Melia}, \&
  {Agol}}]{FalckeMeliaAgol2000}
{Falcke}, H., {Melia}, F., \& {Agol}, E. 2000, \apjl, 528, L13

\bibitem[{{Ferrari}(1998)}]{Ferrari1998}
{Ferrari}, A. 1998, \araa, 36, 539

\bibitem[{{Fish} {et~al.}(2013){Fish}, {Alef}, {Anderson}, {Asada}, {Baudry},
  {Broderick}, {Carilli}, {Colomer}, {Conway}, {Dexter}, {Doeleman}, {Eatough},
  {Falcke}, {Frey}, {Gabányi}, {Gálvan-Madrid}, {Gammie}, {Giroletti},
  {Goddi}, {Gómez}, {Hada}, {Hecht}, {Honma}, {Humphreys}, {Impellizzeri},
  {Johannsen}, {Jorstad}, {Kino}, {Körding}, {Kramer}, {Krichbaum},
  {Kudryavtseva}, {Laing}, {Lazio}, {Loeb}, {Lu}, {Maccarone}, {Marscher},
  {Mart'{\i}-Vidal}, {Martins}, {Matthews}, {Menten}, {Miller}, {Miller-Jones},
  {Mirabel}, {Muller}, {Nagai}, {Nagar}, {Nakamura}, {Paragi}, {Pradel},
  {Psaltis}, {Ransom}, {Rodr{\'{\i}}guez}, {Rottmann}, {Rushton}, {Shen},
  {Smith}, {Stappers}, {Takahashi}, {Tarchi}, {Tilanus}, {Verbiest},
  {Vlemmings}, {Walker}, {Wardle}, {Wiik}, {Zackrisson}, \&
  {Zensus}}]{FishAlefAnderson2013a}
{Fish}, V., {Alef}, W., {Anderson}, J., {et~al.} 2013, ArXiv e-prints,
  1309.3519

\bibitem[{{Fish} {et~al.}(2016){Fish}, {Johnson}, {Doeleman}, {Broderick},
  {Psaltis}, {Lu}, {Akiyama}, {Alef}, {Algaba}, {Asada}, {Beaudoin},
  {Bertarini}, {Blackburn}, {Blundell}, {Bower}, {Brinkerink}, {Cappallo},
  {Chael}, {Chamberlin}, {Chan}, {Crew}, {Dexter}, {Dexter}, {Dzib}, {Falcke},
  {Freund}, {Friberg}, {Greer}, {Gurwell}, {Ho}, {Honma}, {Inoue}, {Johannsen},
  {Kim}, {Krichbaum}, {Lamb}, {León-Tavares}, {Loeb}, {Loinard}, {MacMahon},
  {Marrone}, {Moran}, {Mościbrodzka}, {Ortiz-León}, {Oyama}, {Özel},
  {Plambeck}, {Pradel}, {Primiani}, {Rogers}, {Rosenfeld}, {Rottmann}, {Roy},
  {Ruszczyk}, {Smythe}, {SooHoo}, {Spilker}, {Stone}, {Strittmatter},
  {Tilanus}, {Titus}, {Vertatschitsch}, {Wagner}, {Wardle}, {Weintroub},
  {Woody}, {Wright}, {Yamaguchi}, {Young}, {Young}, {Zensus}, \&
  {Ziurys}}]{FishJohnsonDoeleman2016a}
{Fish}, V.~L., {Johnson}, M.~D., {Doeleman}, S.~S., {et~al.} 2016, \apj, 820,
  90

\bibitem[{{Genzel} {et~al.}(2010){Genzel}, {Eisenhauer}, \&
  {Gillessen}}]{GenzelEisenhauerGillessen2010a}
{Genzel}, R., {Eisenhauer}, F., \& {Gillessen}, S. 2010, Reviews of Modern
  Physics, 82, 3121

\bibitem[{{Ghez} {et~al.}(1998){Ghez}, {Klein}, {Morris}, \&
  {Becklin}}]{GhezKleinMorris1998}
{Ghez}, A.~M., {Klein}, B.~L., {Morris}, M., \& {Becklin}, E.~E. 1998, \apj,
  509, 678

\bibitem[{{Gillessen} {et~al.}(2017){Gillessen}, {Plewa}, {Eisenhauer}, {Sari},
  {Waisberg}, {Habibi}, {Pfuhl}, {George}, {Dexter}, {von Fellenberg}, {Ott},
  \& {Genzel}}]{GillessenPlewaEisenhauer2017a}
{Gillessen}, S., {Plewa}, P.~M., {Eisenhauer}, F., {et~al.} 2017, \apj, 837, 30

\bibitem[{{Goddi} {et~al.}(2017){Goddi}, {Falcke}, {Kramer}, {Rezzolla},
  {Brinkerink}, {Bronzwaer}, {Davelaar}, {Deane}, {de Laurentis}, {Desvignes},
  {Eatough}, {Eisenhauer}, {Fraga-Encinas}, {Fromm}, {Gillessen}, {Grenzebach},
  {Issaoun}, {Janßen}, {Konoplya}, {Krichbaum}, {Laing}, {Liu}, {Lu},
  {Mizuno}, {Moscibrodzka}, {Müller}, {Olivares}, {Pfuhl}, {Porth}, {Roelofs},
  {Ros}, {Schuster}, {Tilanus}, {Torne}, {van Bemmel}, {van Langevelde}, {Wex},
  {Younsi}, \& {Zhidenko}}]{GoddiFalckeKramer2017a}
{Goddi}, C., {Falcke}, H., {Kramer}, M., {et~al.} 2017, International Journal
  of Modern Physics D, 26, 1730001

\bibitem[{{Gravity Collaboration} {et~al.}(2017){Gravity Collaboration},
  {Abuter}, {Accardo}, {Amorim}, {Anugu}, {Ávila}, {Azouaoui}, {Benisty},
  {Berger}, {Blind}, {Bonnet}, {Bourget}, {Brandner}, {Brast}, {Buron},
  {Burtscher}, {Cassaing}, {Chapron}, {Choquet}, {Clénet}, {Collin}, {Coudé
  Du Foresto}, {de Wit}, {de Zeeuw}, {Deen}, {Delplancke-Ströbele}, {Dembet},
  {Derie}, {Dexter}, {Duvert}, {Ebert}, {Eckart}, {Eisenhauer}, {Esselborn},
  {Fédou}, {Finger}, {Garcia}, {Garcia Dabo}, {Garcia Lopez}, {Gendron},
  {Genzel}, {Gillessen}, {Gonte}, {Gordo}, {Grould}, {Grözinger}, {Guieu},
  {Haguenauer}, {Hans}, {Haubois}, {Haug}, {Haussmann}, {Henning}, {Hippler},
  {Horrobin}, {Huber}, {Hubert}, {Hubin}, {Hummel}, {Jakob}, {Janssen},
  {Jochum}, {Jocou}, {Kaufer}, {Kellner}, {Kendrew}, {Kern}, {Kervella},
  {Kiekebusch}, {Klein}, {Kok}, {Kolb}, {Kulas}, {Lacour}, {Lapeyrère},
  {Lazareff}, {Le Bouquin}, {Lèna}, {Lenzen}, {Lévêque}, {Lippa}, {Magnard},
  {Mehrgan}, {Mellein}, {Mérand}, {Moreno-Ventas}, {Moulin}, {Müller},
  {Müller}, {Neumann}, {Oberti}, {Ott}, {Pallanca}, {Panduro}, {Pasquini},
  {Paumard}, {Percheron}, {Perraut}, {Perrin}, {Pflüger}, {Pfuhl}, {Phan Duc},
  {Plewa}, {Popovic}, {Rabien}, {Ram{\'{\i}}rez}, {Ramos}, {Rau}, {Riquelme},
  {Rohloff}, {Rousset}, {Sanchez-Bermudez}, {Scheithauer}, {Schöller},
  {Schuhler}, {Spyromilio}, {Straubmeier}, {Sturm}, {Suarez}, {Tristram},
  {Ventura}, {Vincent}, {Waisberg}, {Wank}, {Weber}, {Wieprecht}, {Wiest},
  {Wiezorrek}, {Wittkowski}, {Woillez}, {Wolff}, {Yazici}, {Ziegler}, \&
  {Zins}}]{Gravity-CollaborationAbuterAccardo2017a}
{Gravity Collaboration}, {Abuter}, R., {Accardo}, M., {et~al.} 2017, \aap, 602,
  A94

\bibitem[{{Johannsen}(2016)}]{Johannsen2016b}
{Johannsen}, T. 2016, Classical and Quantum Gravity, 33, 124001

\bibitem[{{Johnson} {et~al.}(2015){Johnson}, {Fish}, {Doeleman}, {Marrone},
  {Plambeck}, {Wardle}, {Akiyama}, {Asada}, {Beaudoin}, {Blackburn},
  {Blundell}, {Bower}, {Brinkerink}, {Broderick}, {Cappallo}, {Chael}, {Crew},
  {Dexter}, {Dexter}, {Freund}, {Friberg}, {Gold}, {Gurwell}, {Ho}, {Honma},
  {Inoue}, {Kosowsky}, {Krichbaum}, {Lamb}, {Loeb}, {Lu}, {MacMahon},
  {McKinney}, {Moran}, {Narayan}, {Primiani}, {Psaltis}, {Rogers}, {Rosenfeld},
  {SooHoo}, {Tilanus}, {Titus}, {Vertatschitsch}, {Weintroub}, {Wright},
  {Young}, {Zensus}, \& {Ziurys}}]{JohnsonFishDoeleman2015a}
{Johnson}, M.~D., {Fish}, V.~L., {Doeleman}, S.~S., {et~al.} 2015, Science,
  350, 1242

\bibitem[{{Konoplya} {et~al.}(2016){Konoplya}, {Rezzolla}, \&
  {Zhidenko}}]{KonoplyaRezzollaZhidenko2016a}
{Konoplya}, R., {Rezzolla}, L., \& {Zhidenko}, A. 2016, \prd, 93, 064015

\bibitem[{{Krichbaum} {et~al.}(1998){Krichbaum}, {Graham}, {Witzel}, {Greve},
  {Wink}, {Grewing}, {Colomer}, {de Vicente}, {Gomez-Gonzalez}, {Baudry}, \&
  {Zensus}}]{KrichbaumGrahamWitzel1998a}
{Krichbaum}, T.~P., {Graham}, D.~A., {Witzel}, A., {et~al.} 1998, \aap, 335,
  L106

\bibitem[{{Krichbaum} {et~al.}(2014){Krichbaum}, {Roy}, {Lu}, {Zensus}, {Fish},
  {Doeleman}, \& {Event Horizon Telescope (EHT)
  Collaboration}}]{KrichbaumRoyLu2014a}
{Krichbaum}, T.~P., {Roy}, A., {Lu}, R.-S., {et~al.} 2014, in Proceedings of
  the 12th European VLBI Network Symposium and Users Meeting (EVN 2014). 7-10
  October 2014. Cagliari, Italy., 13

\bibitem[{{Lu} {et~al.}(2016){Lu}, {Roelofs}, {Fish}, {Shiokawa}, {Doeleman},
  {Gammie}, {Falcke}, {Krichbaum}, \& {Zensus}}]{LuRoelofsFish2016a}
{Lu}, R.-S., {Roelofs}, F., {Fish}, V.~L., {et~al.} 2016, \apj, 817, 173

\bibitem[{{Luminet}(1979)}]{Luminet1979a}
{Luminet}, J.-P. 1979, \aap, 75, 228

\bibitem[{{Lynden-Bell} \& {Rees}(1971)}]{Lynden-BellRees1971}
{Lynden-Bell}, D. \& {Rees}, M.~J. 1971, \mnras, 152, 461

\bibitem[{{Marrone} {et~al.}(2007){Marrone}, {Moran}, {Zhao}, \&
  {Rao}}]{MarroneMoranZhao2007}
{Marrone}, D.~P., {Moran}, J.~M., {Zhao}, J.-H., \& {Rao}, R. 2007, \apjl, 654,
  L57

\bibitem[{{Matsushita} {et~al.}(2015){Matsushita}, {Asada}, {Blundell},
  {Chang}, {Chen}, {Grimes}, {Han}, {Hirashita}, {Ho}, {Huang}, {Inoue},
  {Jiang}, {Koch}, {Kubo}, {Martin-Cocher}, {Meyer-Zhao}, {Nakamura},
  {Nishioka}, {Nystrom}, {Paine}, {Patel}, {Pu}, {Raffin}, {Snow}, \&
  {Srinivasan}}]{MatsushitaAsadaBlundell2015a}
{Matsushita}, S., {Asada}, K., {Blundell}, R., {et~al.} 2015, IAU General
  Assembly, 22, 2251138

\bibitem[{{Matthews} {et~al.}(2017){Matthews}, {Crew}, {Doeleman}, {Lacasse},
  {Saez}, {Alef}, {Akiyama}, {Amestica}, {Anderson}, {Barkats}, {Baudry},
  {Brogiere}, {Escoffier}, {Fish}, {Greenberg}, {Hecht}, {Hiriart}, {Hirota},
  {Honma}, {Ho}, {Impellizzeri}, {Inoue}, {Kohno}, {Lopez}, {Marti-Vidal},
  {Messias}, {Meyer-Zhao}, {Mora-Klein}, {Nagar}, {Nishioka}, {Oyama},
  {Pankratius}, {Perez}, {Phillips}, {Pradel}, {Rottmann}, {Roy}, {Ruszczyk},
  {Shillue}, {Suzuki}, \& {Treacy}}]{MatthewsCrewDoeleman2017a}
{Matthews}, L.~D., {Crew}, G.~B., {Doeleman}, S.~S., {et~al.} 2017, ArXiv
  e-prints

\bibitem[{{Melia} \& {Falcke}(2001)}]{MeliaFalcke2001a}
{Melia}, F. \& {Falcke}, H. 2001, \araa, 39, 309

\bibitem[{{Mezger} {et~al.}(1989){Mezger}, {Zylka}, {Salter}, {Wink}, {Chini},
  {Kreysa}, \& {Tuffs}}]{MezgerZylkaSalter1989}
{Mezger}, P.~G., {Zylka}, R., {Salter}, C.~J., {et~al.} 1989, \aap, 209, 337

\bibitem[{{Mizuno} {et~al.}(2018){Mizuno}, {Younsi}, {Fromm}, Porth, {de
  Laurentis}, Olivares, {Falcke}, {Kramer}, \& {Rezzolla}}]{MizunoDilaton2018}
{Mizuno}, Y., {Younsi}, Z., {Fromm}, C.~M., {et~al.} 2018, tbd., to be
  submitted

\bibitem[{{Moscibrodzka}(2017)}]{Moscibrodzka2017a}
{Moscibrodzka}, M. 2017, in IAU Symposium, Vol. 322, The Multi-Messenger
  Astrophysics of the Galactic Centre, ed. R.~M. {Crocker}, S.~N. {Longmore},
  \& G.~V. {Bicknell}, 43--49

\bibitem[{{Moscibrodzka} \& {Falcke}(2013)}]{MoscibrodzkaFalcke2013a}
{Moscibrodzka}, M. \& {Falcke}, H. 2013, \aap, 559, L3

\bibitem[{{Moscibrodzka} {et~al.}(2014){Moscibrodzka}, {Falcke}, {Shiokawa}, \&
  {Gammie}}]{MoscibrodzkaFalckeShiokawa2014a}
{Moscibrodzka}, M., {Falcke}, H., {Shiokawa}, H., \& {Gammie}, C.~F. 2014,
  \aap, 570, A7

\bibitem[{{Moscibrodzka} {et~al.}(2009){Moscibrodzka}, {Gammie}, {Dolence},
  {Shiokawa}, \& {Leung}}]{MoscibrodzkaGammieDolence2009}
{Moscibrodzka}, M., {Gammie}, C.~F., {Dolence}, J.~C., {Shiokawa}, H., \&
  {Leung}, P.~K. 2009, \apj, 706, 497

\bibitem[{{Narayan} {et~al.}(1995){Narayan}, {Yi}, \&
  {Mahadevan}}]{NarayanYiMahadevan1995a}
{Narayan}, R., {Yi}, I., \& {Mahadevan}, R. 1995, \nat, 374, 623

\bibitem[{{Novikov} \& {Thorne}(1973)}]{NovikovThorne1973a}
{Novikov}, I.~D. \& {Thorne}, K.~S. 1973, in Black Holes (Les Astres Occlus),
  ed. C.~{Dewitt} \& B.~S. {Dewitt}, 343--450

\bibitem[{{Page} \& {Thorne}(1974)}]{PageThorne1974}
{Page}, D.~N. \& {Thorne}, K.~S. 1974, \apj, 191, 499

\bibitem[{{Porth} {et~al.}(2017){Porth}, {Olivares}, {Mizuno}, {Younsi},
  {Rezzolla}, {Moscibrodzka}, {Falcke}, \& {Kramer}}]{PorthOlivaresMizuno2017a}
{Porth}, O., {Olivares}, H., {Mizuno}, Y., {et~al.} 2017, Computational
  Astrophysics and Cosmology, 4, 1

\bibitem[{{Psaltis} {et~al.}(2016){Psaltis}, {Wex}, \&
  {Kramer}}]{PsaltisWexKramer2016a}
{Psaltis}, D., {Wex}, N., \& {Kramer}, M. 2016, \apj, 818, 121

\bibitem[{{Psaltis} {et~al.}(2015){Psaltis}, {Özel}, {Chan}, \&
  {Marrone}}]{PsaltisOzelChan2015a}
{Psaltis}, D., {Özel}, F., {Chan}, C.-K., \& {Marrone}, D.~P. 2015, \apj, 814,
  115

\bibitem[{{Reid} \& {Brunthaler}(2004)}]{ReidBrunthaler2004a}
{Reid}, M.~J. \& {Brunthaler}, A. 2004, \apj, 616, 872

\bibitem[{{Roelofs} {et~al.}(2017){Roelofs}, {Falcke}, {Brinkerink},
  {Moscibrodzka}, {Gurvits}, Martin-Neira, Kudriashov, Kramer, \&
  {Rezzolla}}]{RoelofsEHI2017}
{Roelofs}, F., {Falcke}, H., {Brinkerink}, C., {et~al.} 2017, \aap, {submitted}

\bibitem[{{Shen} {et~al.}(2005){Shen}, {Lo}, {Liang}, {Ho}, \&
  {Zhao}}]{ShenLoLiang2005}
{Shen}, Z.-Q., {Lo}, K.~Y., {Liang}, M.-C., {Ho}, P.~T.~P., \& {Zhao}, J.-H.
  2005, \nat, 438, 62

\bibitem[{{Sulentic} {et~al.}(2012){Sulentic}, {Marziani}, \&
  {D'Onofrio}}]{SulenticMarzianiDOnofrio2012a}
{Sulentic}, J.~W., {Marziani}, P., \& {D'Onofrio}, M. 2012, in Astrophysics and
  Space Science Library, Vol. 386, Fifty Years of Quasars: From Early
  Observations and Ideas to Future Research, ed. M.~{D'Onofrio}, P.~{Marziani},
  \& J.~W. {Sulentic}, 549

\bibitem[{{Tilanus} {et~al.}(2014){Tilanus}, {Krichbaum}, {Zensus}, {Baudry},
  {Bremer}, {Falcke}, {Giovannini}, {Laing}, {van Langevelde}, {Vlemmings},
  {Abraham}, {Afonso}, {Agudo}, {Alberdi}, {Alcolea}, {Altamirano}, {Asadi},
  {Assaf}, {Augusto}, {Baczko}, {Boeck}, {Boller}, {Bondi}, {Boone}, {Bourda},
  {Brajsa}, {Brand}, {Britzen}, {Bujarrabal}, {Cales}, {Casadio}, {Casasola},
  {Castangia}, {Cernicharo}, {Charlot}, {Chemin}, {Clenet}, {Colomer},
  {Combes}, {Cordes}, {Coriat}, {Cross}, {D'Ammando}, {Dallacasa}, {Desmurs},
  {Eatough}, {Eckart}, {Eisenacher}, {Etoka}, {Felix}, {Fender}, {Ferreira},
  {Freeland}, {Frey}, {Fromm}, {Fuhrmann}, {Gabanyi}, {Galvan-Madrid},
  {Giroletti}, {Goddi}, {Gomez}, {Gourgoulhon}, {Gray}, {di Gregorio},
  {Greimel}, {Grosso}, {Guirado}, {Hada}, {Hanslmeier}, {Henkel}, {Herpin},
  {Hess}, {Hodgson}, {Horns}, {Humphreys}, {Hutawarakorn Kramer}, {Ilyushin},
  {Impellizzeri}, {Ivanov}, {Julião}, {Kadler}, {Kerins}, {Klaassen}, {van 't
  Klooster}, {Kording}, {Kozlov}, {Kramer}, {Kreikenbohm}, {Kurtanidze},
  {Lazio}, {Leite}, {Leitzinger}, {Lepine}, {Levshakov}, {Lico}, {Lindqvist},
  {Liuzzo}, {Lobanov}, {Lucas}, {Mannheim}, {Marcaide}, {Markoff},
  {Mart{\'{\i}}-Vidal}, {Martins}, {Masetti}, {Massardi}, {Menten}, {Messias},
  {Migliari}, {Mignano}, {Miller-Jones}, {Minniti}, {Molaro}, {Molina},
  {Monteiro}, {Moscadelli}, {Mueller}, {Müller}, {Muller}, {Niederhofer},
  {Odert}, {Olofsson}, {Orienti}, {Paladino}, {Panessa}, {Paragi}, {Paumard},
  {Pedrosa}, {Pérez-Torres}, {Perrin}, {Perucho}, {Porquet}, {Prandoni},
  {Ransom}, {Reimers}, {Rejkuba}, {Rezzolla}, {Richards}, {Ros}, {Roy},
  {Rushton}, {Savolainen}, {Schulz}, {Silva}, {Sivakoff}, {Soria-Ruiz},
  {Soria}, {Spaans}, {Spencer}, {Stappers}, {Surcis}, {Tarchi}, {Temmer},
  {Thompson}, {Torrelles}, {Truestedt}, {Tudose}, {Venturi}, {Verbiest},
  {Vieira}, {Vielzeuf}, {Vincent}, {Wex}, {Wiik}, {Wiklind}, {Wilms},
  {Zackrisson}, \& {Zechlin}}]{TilanusKrichbaumZensus2014a}
{Tilanus}, R.~P.~J., {Krichbaum}, T.~P., {Zensus}, J.~A., {et~al.} 2014, ArXiv
  e-prints, 1406.4650

\bibitem[{{Viergutz}(1993)}]{Viergutz1993a}
{Viergutz}, S.~U. 1993, \aap, 272, 355

\bibitem[{{Yuan} {et~al.}(2002){Yuan}, {Markoff}, \&
  {Falcke}}]{YuanMarkoffFalcke2002}
{Yuan}, F., {Markoff}, S., \& {Falcke}, H. 2002, \aap, 383, 854

\bibitem[{{Zensus} \& {Falcke}(1999)}]{ZensusFalcke1999a}
{Zensus}, J.~A. \& {Falcke}, H. 1999, in Astronomical Society of the Pacific
  Conference Series, Vol. 186, The Central Parsecs of the Galaxy, ed.
  H.~{Falcke}, A.~{Cotera}, W.~J. {Duschl}, F.~{Melia}, \& M.~J. {Rieke} (San
  Francisco: Astronomical Society of the Pacific), 118

\end{thebibliography}

\end{document}